\documentclass[article]{WileyASNA-v1}

\newcommand{\lx}{L_{\rm X}}
\newcommand{\lo}{L_{\rm UV}}
\newcommand{\gx}{\Gamma_{\rm X}}
\newcommand{\fvar}{F_{\rm var}}
\newcommand{\sfvar}{\sigma(F_{\rm var})}
\newcommand{\deltat}{\Delta t_{\rm sys}}

\newcommand{\rev}[1]{{ #1}}

\articletype{Original Article}%
\received{\today}
\revised{--}
\accepted{--}

\begin{document}

\title{The non-linear X-ray/UV relation in active galactic nuclei: contribution of instrumental effects on the X-ray variability}

\author[1]{Elisabeta Lusso*}

\authormark{Elisabeta Lusso} % \textsc{et al}}
\address[1]{\orgdiv{Centre for Extragalactic Astronomy}, \orgname{Durham University}, \orgaddress{\state{South Road, Durham, DH1 3LE}, \country{UK}}}

\corres{*\email{elisabeta.lusso@durham.ac.uk}}

\abstract{We have recently demonstrated that the non-linear relation between ultraviolet and X-ray luminosity in quasars is very tight (with an intrisic dispersion of $\sim$0.2 dex), once contaminants (e.g. dust reddening, X-ray absorption), variability, and differences in the active galactic nuclei (AGN) physical properties are taken into account. This relation has thus the great potential to advance our understanding in both supermassive black hole accretion physics and observational cosmology, by targeting a single class of objects. 
Here we focus on the various contributions to the {\it observed} X-ray variability in a homogenous sample of 791 quasars selected from SDSS--DR7 with X-ray data from the 3XMM--DR7 source catalogue. The 250 quasars in this cleaned data set with at least two X-ray observations typically vary with a standard deviation of fractional variation of 15--30\% on timescales of weeks/years. Yet, when the count rates are computed at progressively smaller off-axis values, the same quantity is reduced to roughly 10--25\%. This suggests that, when estimating variability indicators, part of the quoted variability amplitude could be due to instrumental/calibration issues rather than {\it true} variations in the quasar emission.}


% WARNING: it appears twice in the pdf! 
\jnlcitation{\cname{%
\author{Lusso E.}} (\cyear{2018}), 
\ctitle{The non-linear X-ray/UV relation in active galactic nuclei: contribution of instrumental effects on the X-ray variability}, \cjournal{Astron. Nachr.}, \cvol{--}.}

\fundingInfo{}

\maketitle

%\footnotetext{\textbf{Abbreviations:} ANA, anti-nuclear antibodies; APC, antigen-presenting cells; IRF, interferon regulatory factor}

\section{Introduction}\label{intro}
X-ray variability is a powerful tool to probe both the nuclear region and the central engine in AGN. Extensive observational campaigns of single objects (e.g. \citealt{uttley2003,2018arXiv180310444A}) have revealed that this high-energy variability is mostly a stochastic process involving very different timescales (e.g. from hours to weeks/years), which does not show a clear dependence on environmental effects that could trigger and enhance variability (e.g. enhanced accretion processes or dynamical instabilities; \citealt{paolillo2017}). %Despite the significant telescope time invested in studying this phenomenon, the physical mechanism that causes the X-ray variability is still poorly known. 
Several mechanisms to explain the observed spectral variations and/or inter-band time lags have been proposed, such as {\it internal} processes occurring in the immediate vicinity of the black hole (e.g. accretion instabilities, coronal flaring; \citealt{2006MNRAS.367..801A,goosmann2006}) or {\it external} at larger scales (e.g. broad line region, variable cloud occultations; \citealt{nardinirisaliti2011,2016A&A...596A..79S}). 

Moreover, monitoring campaigns over years/decades have shown that the X-ray variability is somewhat correlated with the AGN physical parameters, such as the black hole mass ($M_{\rm BH}$), the accretion rate (parametrized by the Eddington ratio between bolometric and Eddington luminosities, $\lambda_{\rm Edd}=L_{\rm bol}/L_{\rm Edd}$) and the nuclear X-ray luminosity. The observed anti-correlation between the normalized excess variance $\sigma_{\rm rms}^2$ (a proxy of the X-ray variability amplitude; \citealt{vaughan2003}) and the X-ray luminosity in a given band implies that the higher the AGN luminosity the lower its {\it intrinsic} variability (e.g. \citealt{young2012,shemmer2014,lanzuisi2014,yang2016,paolillo2017,zheng2017}). Various models have been proposed in the literature to interpret this anti-correlation. Some invoke a characteristic timescale of the accretion disc (e.g. the viscous timescale), or a physical timescale directly linked to the X-ray emission mechanism, i.e. the cooling time of the electrons in the Comptonisation process (e.g. see \citealt{2012A&A...540L...2I} and references therein). 

%it could be interpreted as due to the fact that high luminous AGN are usually observed with high black hole mass value where the last stable orbit is at larger radii, which, in turn, somewhat suppress the power of X-ray variability. 

Regardless of the specific details of these models, all of them point toward a scenario where $M_{\rm BH}$ is the main driver of the correlation, rather than the luminosity. 
X-ray variability thus becomes a tool to provide an independent estimate of $M_{\rm BH}$ (e.g. \citealt{hayashida1998,luyu2001,czerny2001,papadakis2004,uttleymch2005,2005ApJ...625L..39M,mch2006,uttley2007,ponti2012,demarco2013,2018arXiv180306891M}). 

Decades of studies and surveys with several Ms invested from all the major space observatories have provided us with exquisite data and good sample statistics, allowing us to expand our understanding of the connection between X-ray variability and accretion properties. Yet, \rev{the physical mechanisms causing the X-ray variability in AGN still remains an open question}. The observed correlations between excess variance and AGN physical parameters suffer from a large scatter ($\geq 0.4$ dex), where calibration effects may also contribute to the observed dispersion. \rev{\cite{zhou2010} have indeed shown that, once calibrated, the correlation between the X-ray variability amplitude and $M_{\rm BH}$ for 21 reverberation-mapped AGN appears very tight, with an intrinsic dispersion of only 0.20 dex.} %hampering both the determination of the intrinsic distribution of these parameters through variability, and any possible discriminations amongst the various models. 

\rev{Moreover, the non-linear relation between the X-ray (parametrized by the rest-frame monochromatic luminosity/flux at 2 keV, $\lx$) and the ultraviolet (at 2500~\AA, $\lo$) emission in AGN also appears to be much tighter than previously thought once contaminants are taken into account \citep{lusso16}. X-ray variability is a major source of dispersion in the $\lx$--$\lo$ relation, contributing by roughly 0.12 dex, or $\sim$30\% (\citealt{lusso16}; see also \citealt{grupe2010,vagnetti2013,chiaraluce2018}).}
The contribution of variability to the non linear $\lx$--$\lo$ correlation can be ascribed to at least two factors. The first one is the \textit{intra-source} dispersion due to variability within the same object, which can occur on both short timescales (days/weeks), due to variations in the inner nuclear regions, and long timescales (months/years), due to changes occurring in the outer accretion disc (see the recent work by \citealt{chiaraluce2018} and references therein). The second factor is the \textit{inter-source} dispersion, arising from the individual physical properties of the different sources. An additional, often neglected, contribution to the dispersion can be associated with instrumental calibration uncertainties, which affect both the UV and the X-rays, although, in the latter case, calibrations issues have potentially a higher impact on the $\lx$--$\lo$ dispersion itself. In fact, every X-ray catalogue is a collection of several observations taken in the most diverse conditions (e.g. the background level may vary even within the same exposure), with a variety of observing modes, and with detectors whose response changes over time (as, for instance, the sensitivity of the ACIS camera of {\it Chandra}, \citealt{plucinsky2018}). In the case of non pointed (off-axis) observations, the decrease of the effective area with increasing field angle (i.e. vignetting) distorts the point spread function, which, in turn, affects the flux measurement \citep[e.g.][]{aschenbach2002}. In this manuscript, we will try to quantify the contribution of instrumental effects on the total observed X-ray variability, and its implications for the dispersion of the $\lx$--$\lo$ correlation. Understanding the importance of calibration issues in increasing the scatter in this relation is key for two main reasons. 

The first one is related to physics. At the time of writing, {\it all} the works in the literature are in very good agreement in their measurement of the slope $\gamma \sim 0.6$, where log\,$\lx$ = $\gamma$\,log\,$\lo$\,$+\beta$ (e.g. \citealt{steffen06,just07,lusso2010,marchese2012}). {\it Why is the same value of the slope consistently found, even across a large redshift range?} The answer to this question would provide fundamental insights on the physical link between the SMBH accretion disc and the ionized plasma of relativistic electrons responsible of the X-ray emission (the so-called {\it corona}; e.g. \citealt{svenssonzdziarski1994}, see also \citealt{lusso2017} and references therein). The details of the physical mechanisms underlying this ``disc--corona coupling'' are still unknown.

The second reason is cosmological. The $\lx$--$\lo$ relation can be used as a distance indicator, allowing the construction of a {\it Hubble diagram} for quasars \citep{risalitilusso2015,risalitilusso2017,bisogni2017}. \rev{With this technique, it is possible to test the cosmological models and to measure the cosmological parameters (within any chosen model) for the first time up to $z\sim6$, but it could potentially be applied at the highest quasar redshifts observed today, i.e. $z=7.54$ \citep{banados2018}}.

A good comprehension of the role of systematics is thus critical to establish the intrinsic scatter of the relation, possibly free of biases, which is key to advance our understanding of both accretion physics and the evolution of the Universe across cosmic time.

\begin{figure}[t]
\centering\includegraphics[width=8cm]{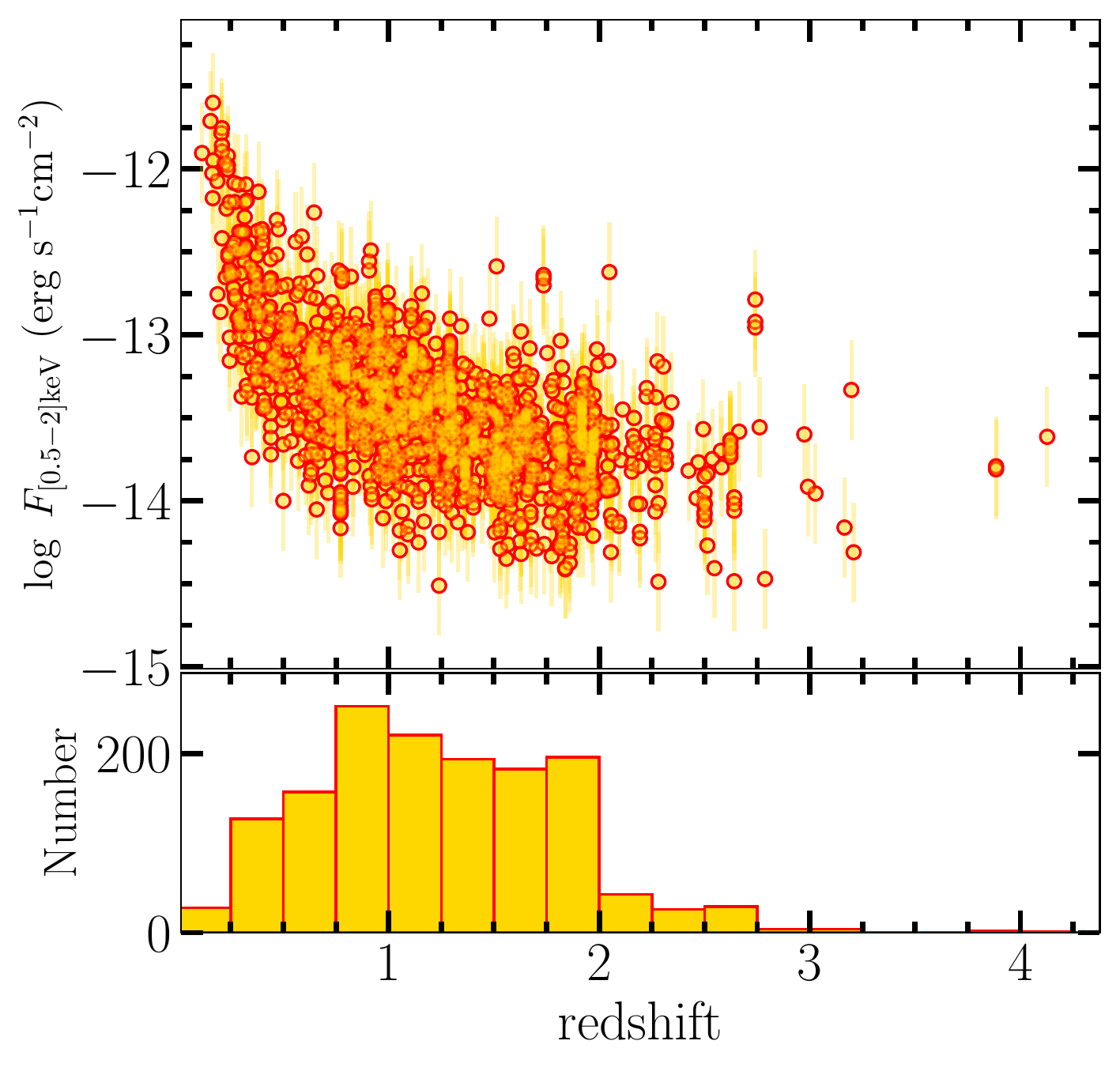}
\caption{Top panel: Distribution of the observed 0.5--2 keV flux for all the observations ($N_{\rm obs}=1,468$) versus redshift for the clean quasar sample (791 unique quasars, 250 of which with $N_{\rm multiple}\geq2$). Bottom panel: Redshift distribution. \label{fig1}}
\end{figure}

\section{The data set}
To define the quasar sample we adopted a similar procedure as described in our previous works \citep{risalitilusso2015,lusso16,lusso2017}. We briefly outline the main steps in the following. The quasar sample is obtained by cross-matching the quasar Sloan Digital Sky Survey catalogue published by \citet{shen2011} with the serendipitous X-ray source catalogue 3XMM--DR7 \citep{rosen2016}. 
For the matching, we adopted a maximum separation of 3$^{\prime\prime}$ to provide optical classification and spectroscopic redshift for all objects. 

To select a {\it clean} sample where biases and contaminants are minimized, we applied a series of filters. Jetted and broad absorption line (BAL) quasars are removed by using the flags in the \citet{shen2011} catalogue, as well as the radio and BAL classification published by \citet{mingo2016} and \citet{gibson2009}, respectively. The following quality cuts from the 3XMM--DR7 catalogue were then applied: SUM\_FLAG$<$3 (low level of spurious detections), HIGH\_BACKGROUND$=$0 (relatively low background levels), and CONFUSED$=$0 (low level of confusion)\footnote{For more details the reader should refer to the 3XMM catalogue user guide at the following website http://xmmssc.irap.omp.eu/Catalogue/3XMM-DR7/3XMM-DR7\_Catalogue\_User\_Guide.html.}. The remaining sample after the cross match is composed by 4,615 {\it XMM-Newton} observations (2,927 unique quasars, 785 of which with 2 or more observations).  

In addition, only sources with low levels of both X-ray absorption (i.e. with an X-ray photon index $\gx>1.7$) and dust reddening ($E(B-V)<0.1$), and with a measurement of both the soft (0.5--2 keV) and hard (2--12 keV) fluxes are considered.
\rev{The {\it Eddington bias} is minimized by including only quasars whose minimum detectable X-ray flux is lower than the expected one in each observation, as estimated from the rest-frame UV emission and assuming a slope of 0.6 (see \citealt{risalitilusso2015,lusso16}).} All the details on this selection procedure, especially the simulations, will be provided in a forthcoming publication. 
The final, clean sample is shown in Figure~\ref{fig1}, and consists of 791 unique quasars (for 1,468 observations) spanning a redshift range of $z = 0.114$--4.127. This is only $\sim$27\% of the initial sample, as a result of the stringent filters discussed above. Of these 791 quasars, 250 objects have multiple ($N_{\rm multiple}\geq 2$) observations \rev{(i.e. 927 observations)}.

\begin{figure}[t]
\centering\includegraphics[width=8.5cm]{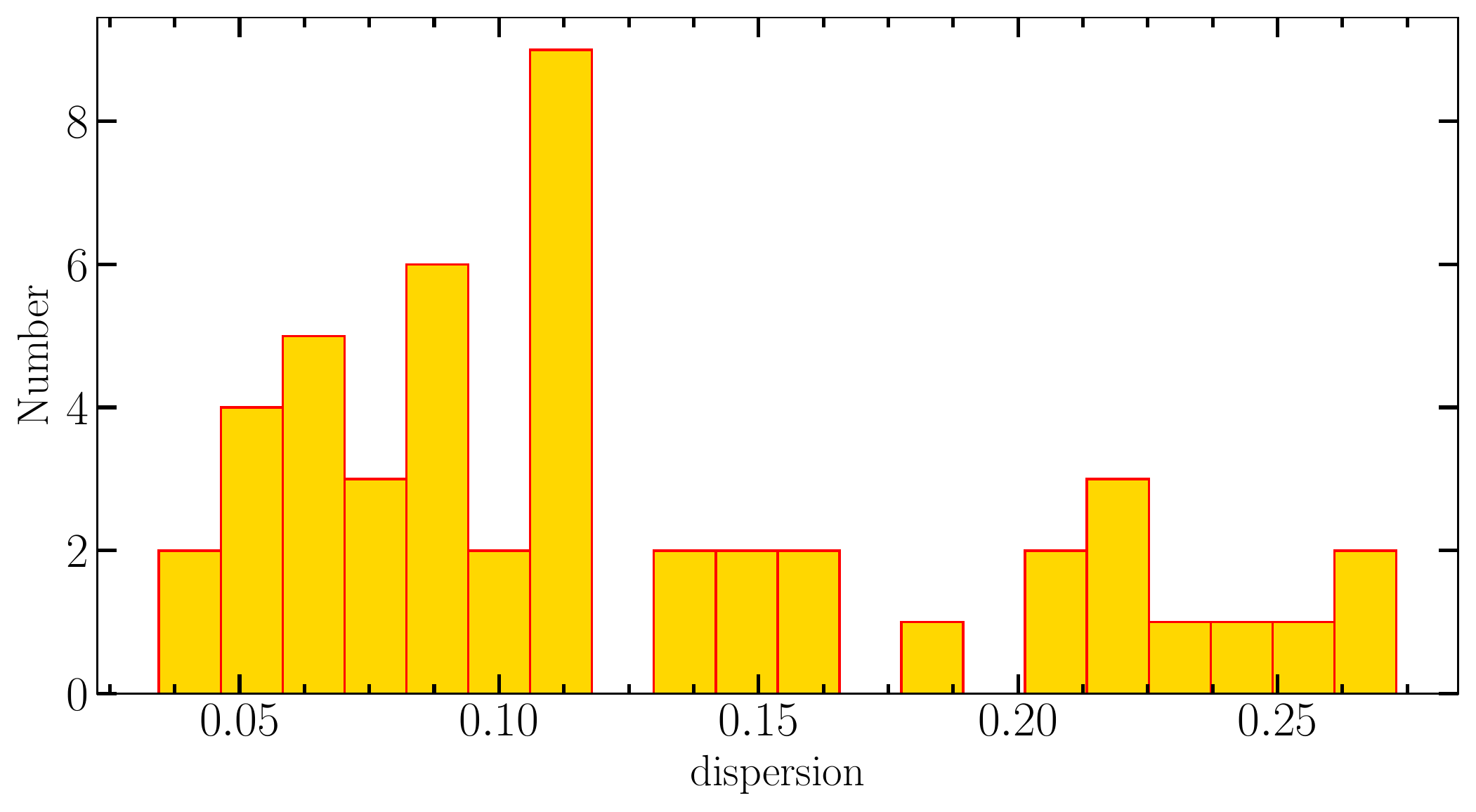}
\caption{Distribution of the dispersion, defined as the 1$\sigma$ standard deviation of the logarithm of the observed 0.5--2 keV fluxes, for the 48 sources with $N_{\rm multiple}\geq5$ (423 observations). The mean (median) value is 0.12 (0.11) dex. \label{figdisp}}
\end{figure}

\section{Dependence of the X-ray flux on the off-axis angle}
In the case of serendipitous observations, as for the majority of the detections listed in the 3XMM catalogues, the X-ray fluxes are often extracted off-axis. We want to investigate any possible variations of the X-ray flux as a function of the off-axis angle in the case of objects observed multiple times. To have reasonable statistics, we restrict the discussion to the objects within the clean sample with 5 or more observations: 48 unique AGN (423 observations in total). 
Figure~\ref{figdisp} shows the distribution of the dispersion, defined as the 1$\sigma$ standard deviation of the logarithm of the observed 0.5--2 keV fluxes, for each of these 48 sources. In other words, this is the distribution of the \textit{intra-source} dispersion of the observed soft fluxes. The range of values is relatively large, from about 10\% to 30\% variations. 
The mean (median) value of this distribution, which is a proxy of the \textit{inter-source} dispersion, is roughly 0.12 (0.11) dex. 

Similar results were found by \citet{lusso16} for the sample of 159 quasars with multiple observations. The majority of these objects have two observations (55\%, 88/159), with only 35 quasars with more than 3 detections. The measured slope ($\gamma$), intercept ($\beta$), and dispersion values of the $\lx$--$\lo$ relation computed by considering the X-ray flux of the longest {\it XMM} observation are $\gamma=0.672\pm0.035$, $\beta=6.044^{+1.166}_{-1.075}$, and 0.23 dex, respectively.
By considering the $\lx$ values estimated as the average between the $1^{\rm st}$ and $2^{\rm nd}$ {\it XMM} longest exposures, the dispersion on the $\lx$--$\lo$ relation reduces to 0.21 dex (see their Figure~6), while both slope and intercept are still in agreement. Overall, the contribution of the \textit{inter-source} dispersion is roughly 0.12 dex.

\begin{figure}
\centering{
\includegraphics[width=7cm]{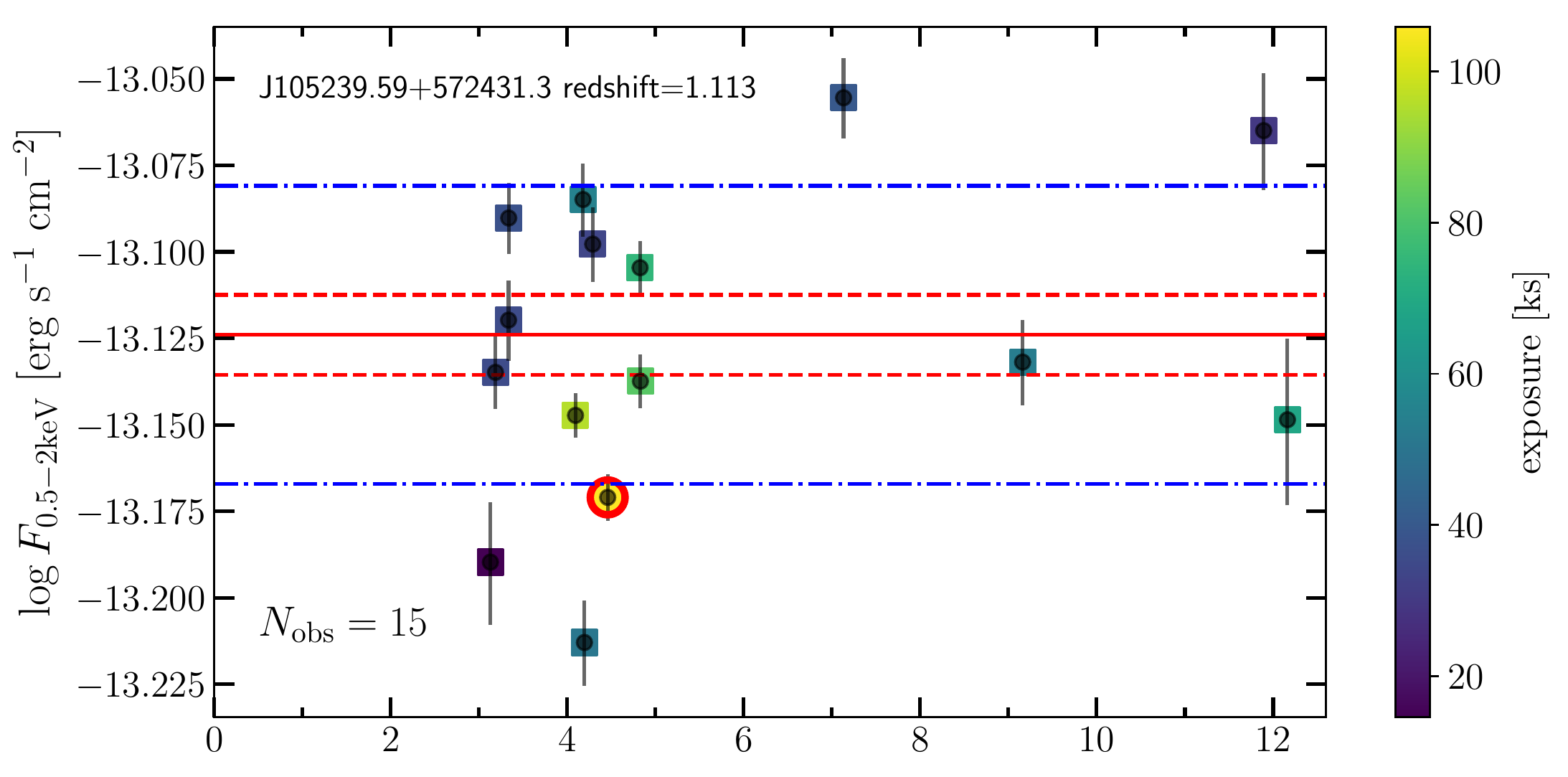}
\includegraphics[width=7cm]{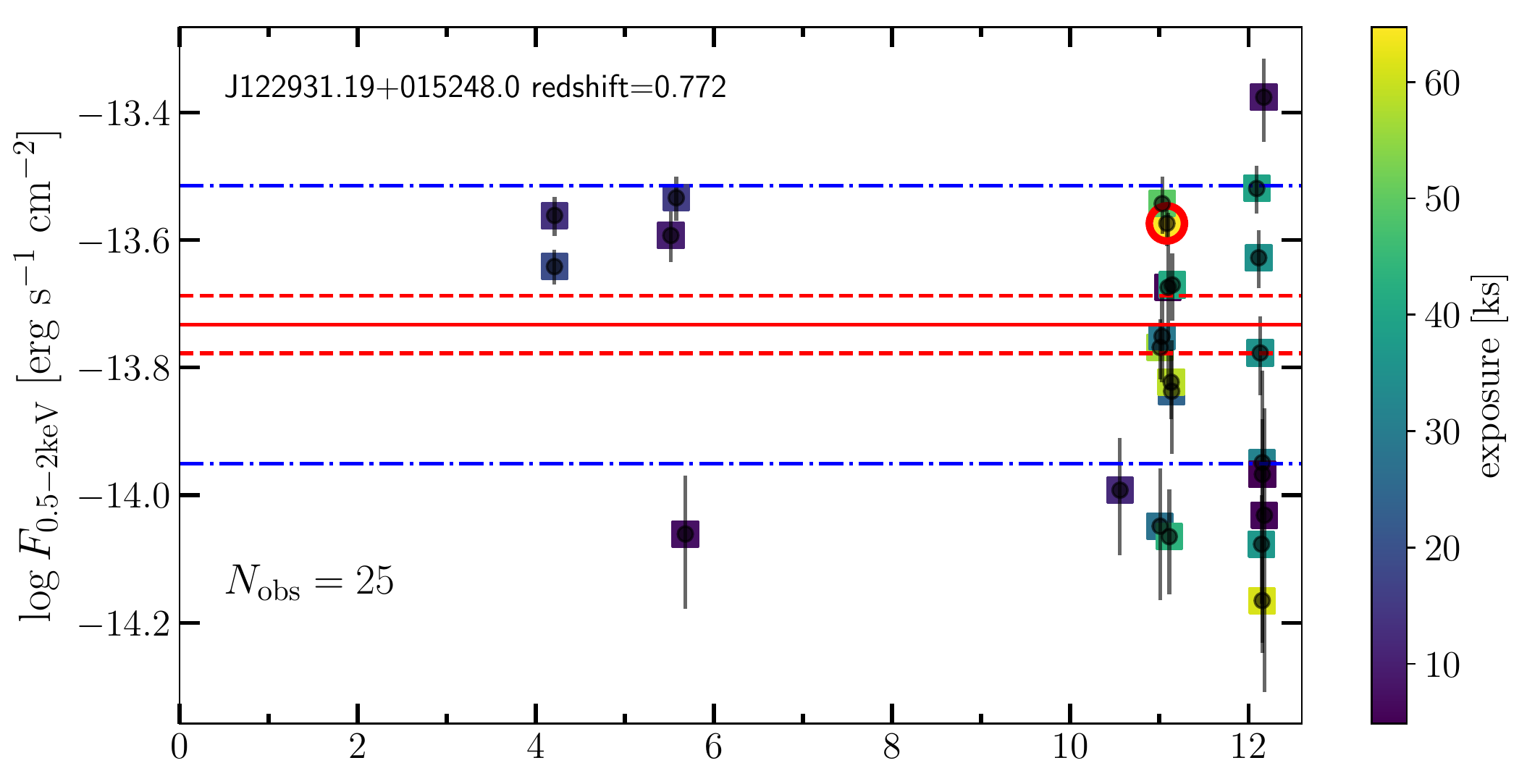}
\includegraphics[width=7cm]{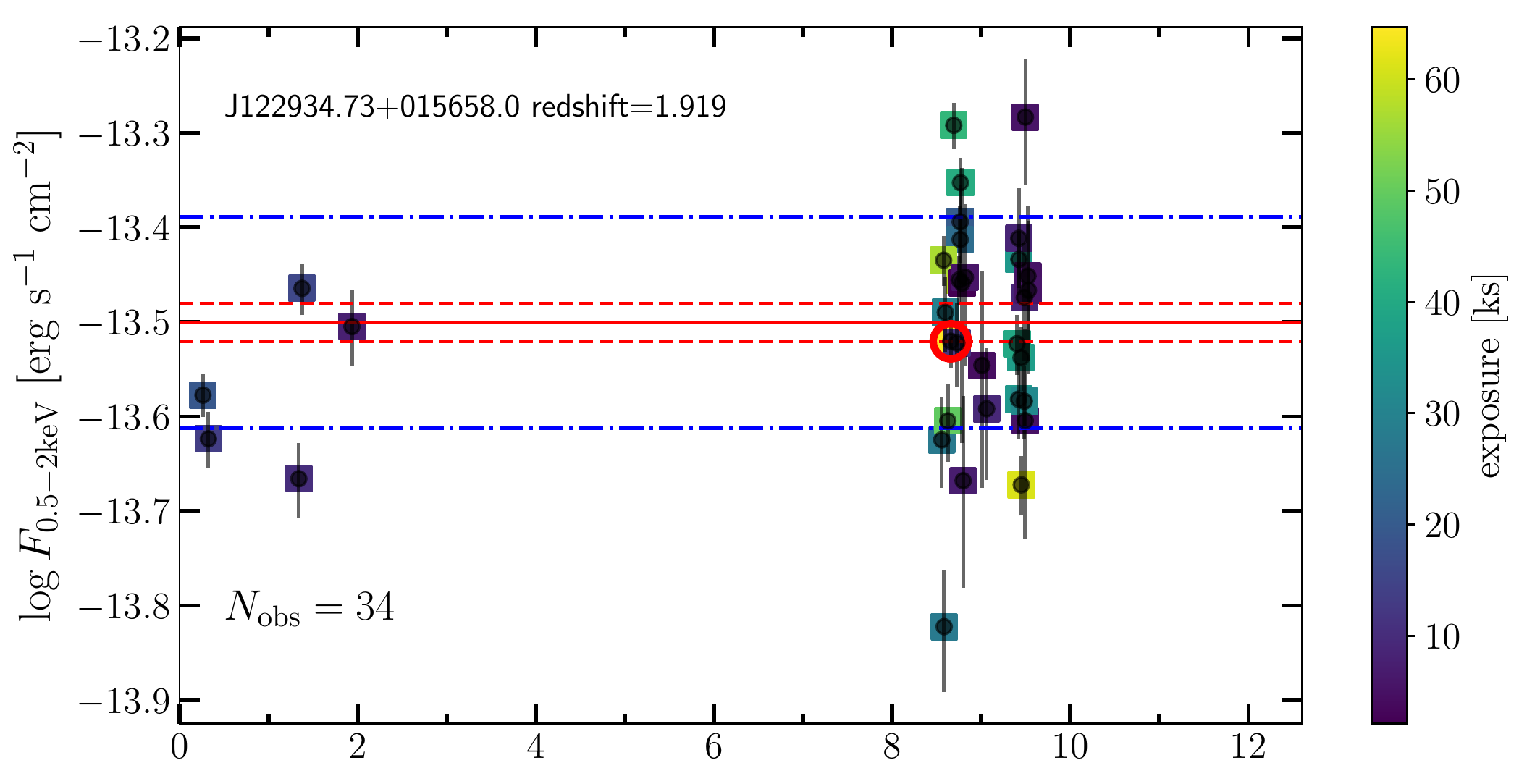}
\includegraphics[width=7cm]{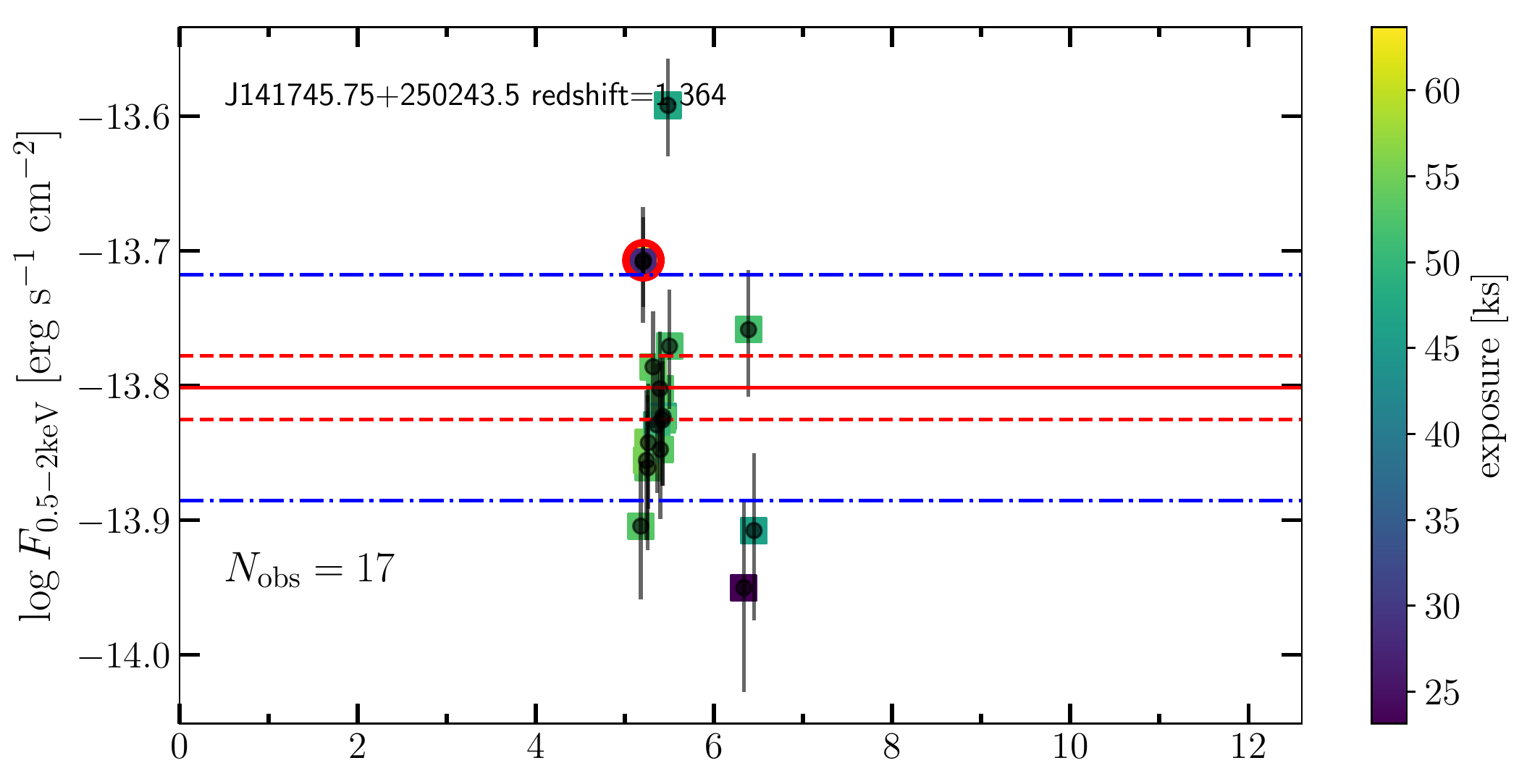}
\includegraphics[width=7cm]{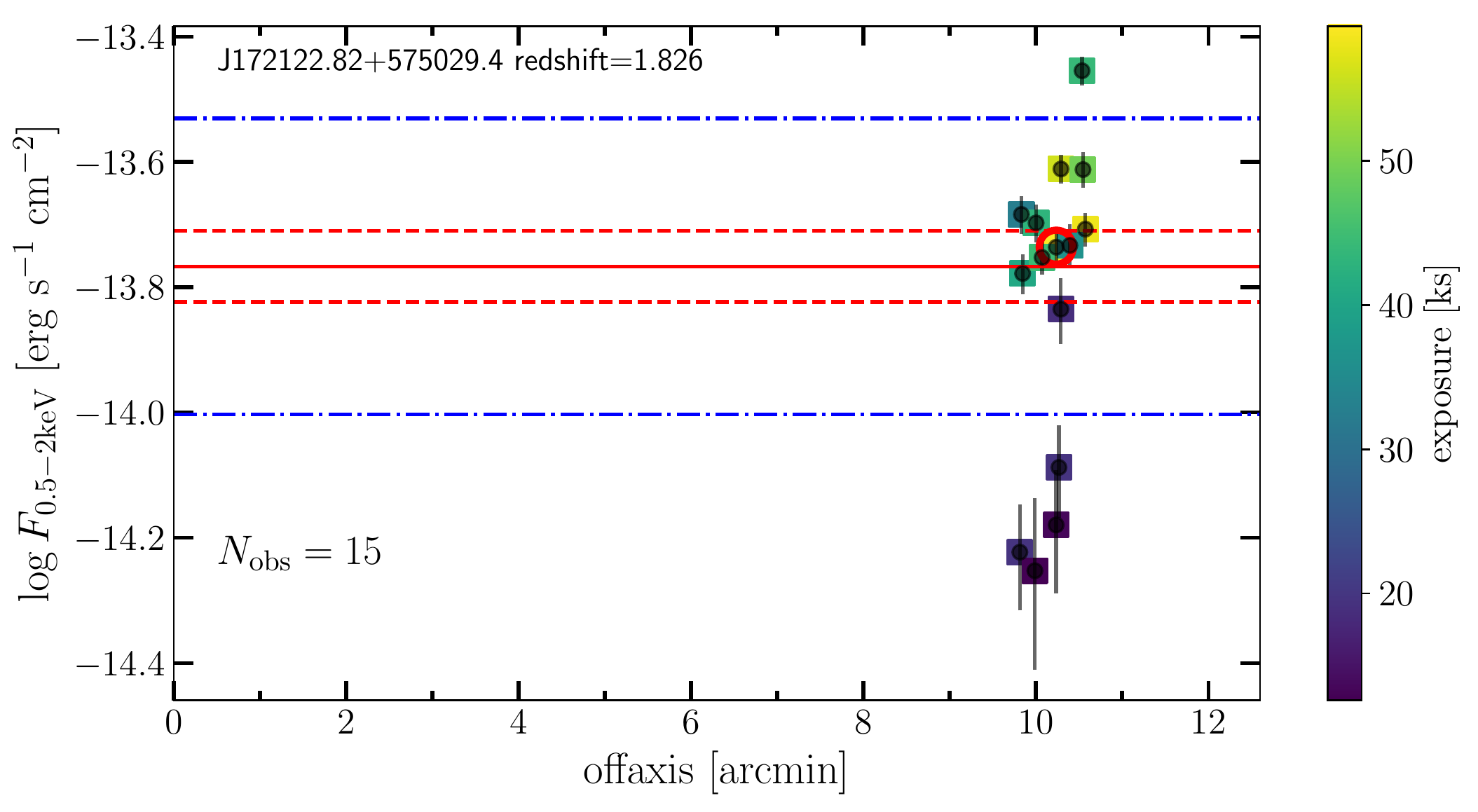}}
\caption{Distribution of the observed soft X-ray flux as a function of the off-axis angle for the 5 AGN fulfilling our filtering criteria and having $N_{\rm multiple}\geq15$. The colour bar represents the EPIC on-time exposure (in ks) for each object. The total number of observations is also reported. The open red circles mark the observations with the longest EPIC exposure. The red solid (dashed) line represents the flux mean (1$\sigma$ uncertainty), while the blue dot-dashed line is the dispersion of the data around the mean. \label{figs}}
\end{figure}

Such a dispersion is a combination of {\it intrinsic} variability and flux variations due to instrumental effects (e.g. background subtraction, vignetting). Disentangling these two factors is not straightforward, as multiple pointed observations performed with similar conditions are challenging (if not impossible) to obtain even for a single object. 
The procedure to convert counts into fluxes (and then luminosities) introduces additional scatter to the intrinsic X-ray variability. 
If we further restrict our sample to the objects with off-axis angles less than 5$^\prime$\footnote{The effective area of the mirrors is a function of off-axis angle. As a result, as the off-axis angle increases, less of the photons entering the telescopes actually reach the focal plane (i.e. the so-called {\it vignetting}). At 5$^\prime$, the vignetting function is around 0.8, meaning that $\sim80$\% of the photons are retained.} (15 AGN, $N_{\rm obs}=62$), we have that the average inter-source dispersion drops to 0.05 dex, with the intra-source values ranging from 0 (on-axis, 4 objects are pointed) to a maximum of 0.12 dex.
Sources with multiple, almost on-axis, observations have uncertainties due to flux calibration, background subtraction, and vignetting correction almost negligible. 

An alternative approach consists in the adoption of the observation with the longest exposure as the best detection measurement, as done in \citet{lusso16,lusso2017}. This choice is reasonable in a statistical sense, yet there could be some cases where the flux of the longest exposure corresponds to a large off-axis angle, thus having the highest calibrations uncertainties. To illustrate this point, in Figure~\ref{figs} we present the distribution of the observed soft X-ray flux as a function of the off-axis angle for the 5 objects that have more that 15 observations. The colour bar represents the EPIC on-time exposure (in ks) for each object. This is the total good time interval of the detector's chip where the source is positioned. The open red circle marks the observation with the longest EPIC exposure. The red solid (dashed) line represents the flux mean (1$\sigma$ uncertainty), while the blue dot-dashed line is the 1$\sigma$ dispersion of the data around the mean.
In almost all cases, the detection with the longest exposure is indeed located at large off-axis angles. Yet, its value is often consistent \rev{within} $\sim$1$\sigma$ compared to the average one. In our future works, we will refine the selection of the {\it best} X-ray observation convolving the net exposure with the off-axis angle.

\begin{figure}[t]
\includegraphics[width=8.5cm]{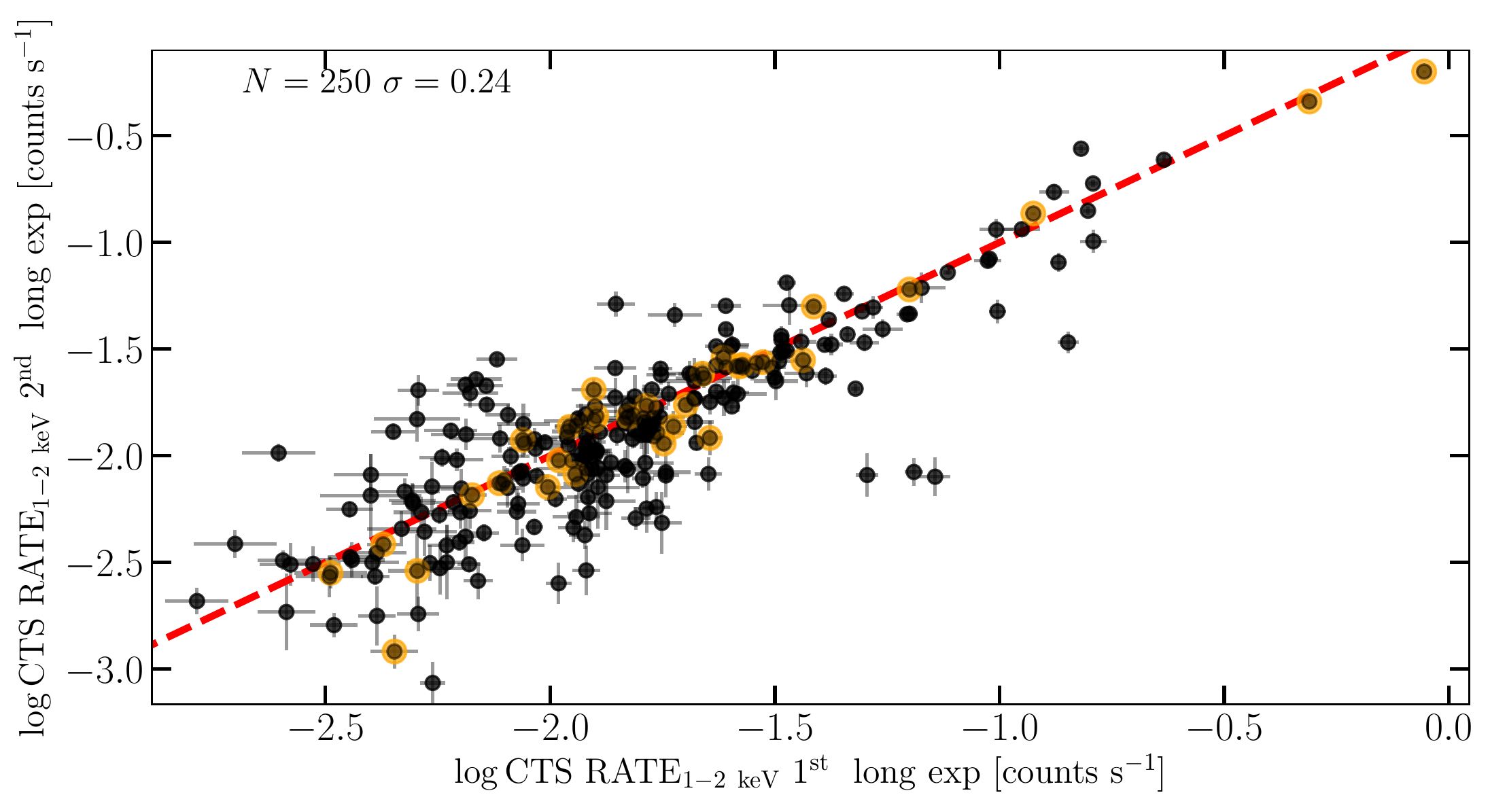}
\caption{Count rates in the 1--2 keV band of the $2^{\rm nd}$ longest {\it XMM} exposure as a function of those with the longest {\it XMM} exposure for the selected quasar sample with multiple observations. The dispersion along the one-to-one relation (red dashed line) is 0.24 dex. Orange points mark a sub-sample of AGN where the count rates are computed at off-axis angles $<5^{\prime}$. The dispersion of this sub-sample drops to 0.15 dex. \label{figcrt}}
\end{figure}

\begin{figure}[t]
\centering{
\includegraphics[width=8cm]{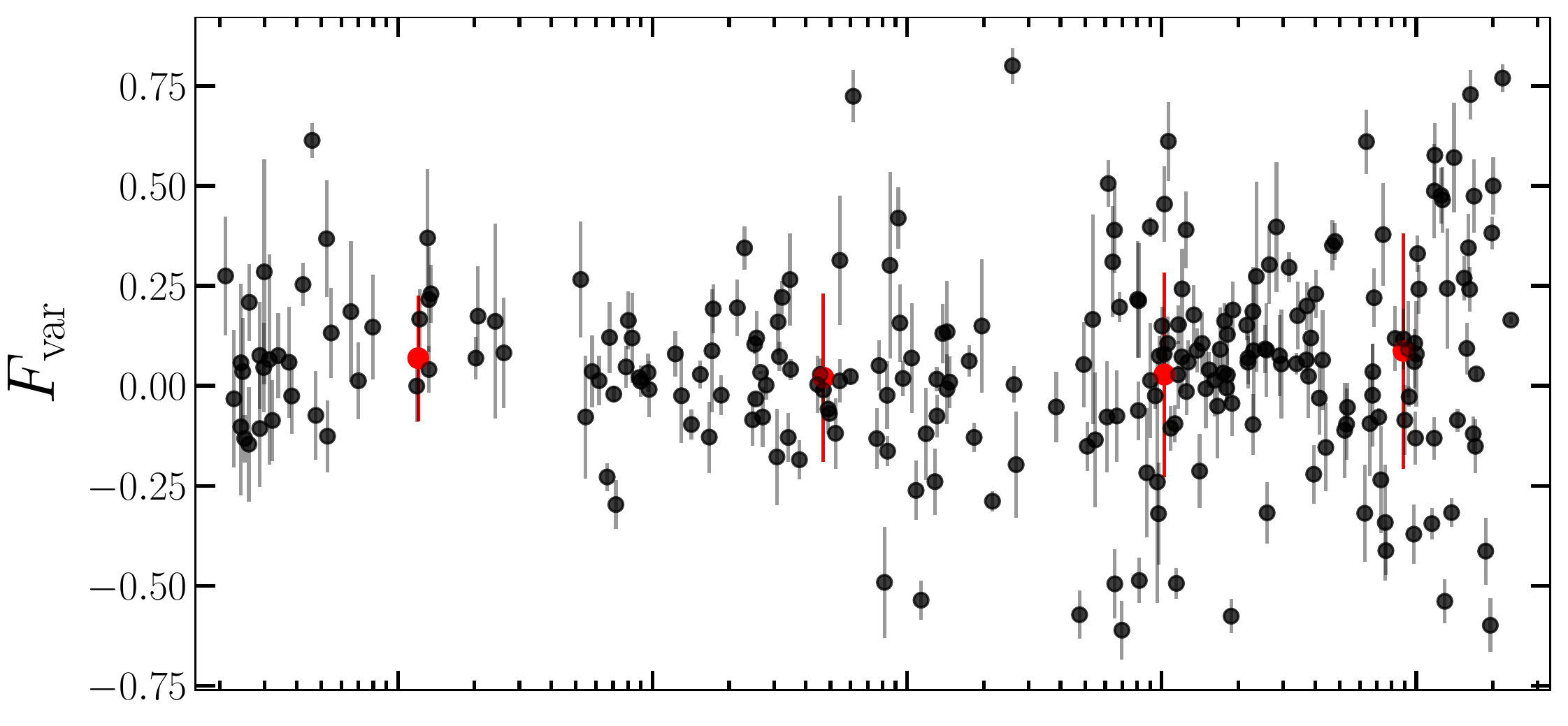}
\includegraphics[width=8cm]{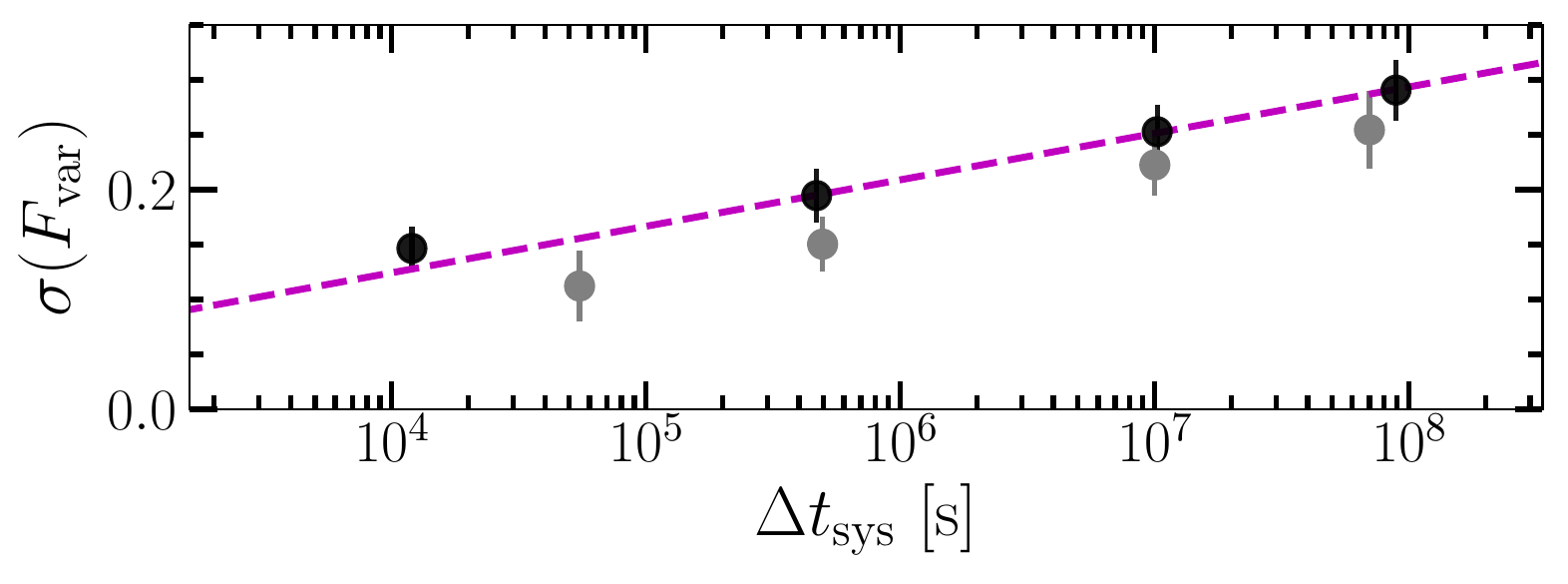}}
\caption{Top: distribution of $\fvar$ as a function of the rest-frame elapsed time $\deltat$. Each point is estimated taking the count rate of the $1^{\rm st}$ and the $2^{\rm nd}$ longest {\it XMM} exposures. The red points represent the mean of $\fvar$ in bins of $\deltat$, whilst the error bars are the 1$\sigma$ dispersion around the mean. Lower panel: Gaussian dispersion of $\fvar$ as a function of $\deltat$. The magenta dashed line is a linear fit of $\sfvar$ (see text for details). \rev{The grey points represent the $\sfvar$ as a function of $\deltat$ for the subsample of 145 quasars with offaxis $<10^{\prime}$}.\label{figvar}}
\end{figure}

\section{Fractional variation}
We can provide another quantitative measure of the amplitude of the total X-ray variability in our sample by following a similar procedure as the one described by \citet{gibson2012} and \citet{lusso16}. We considered the 250 AGN with more than 2 observations and we computed the {\it fractional variation} ($\fvar$) as
\begin{equation}
\label{fvar}
\fvar \equiv (c_i - c_j)/(c_i + c_j),
\end{equation}
where $c_i$ and $c_j$ are the count rates for the $1^{\rm st}$ and $2^{\rm nd}$ {\it XMM} longest exposures, respectively. A comparison of the count rates employed to estimate $\fvar$ is shown in Figure~\ref{figcrt}. The 3XMM-DR7 count rates have been background subtracted and corrected for vignetting and PSF \rev{losses}\footnote{http://xmmssc.irap.omp.eu/Catalogue/3XMM-DR5/col\_srcpar.html}.
Each $\fvar$ measurement between these two exposures is associated with a rest-frame elapsed time $\deltat$ defined as the absolute value $\deltat = \lvert t_j - t_i \rvert / (1+z)$. We then assumed an intrinsic Gaussian distribution for $\fvar$ and estimated the standard deviation of this distribution, $\sfvar$, using the likelihood method described by \citet{maccacaro1988}. We finally binned $\deltat$ in four intervals of about 50 epochs each.
The upper and lower panels of Figure~\ref{figvar} show our results for $\fvar$ and $\sfvar$ as a function of $\deltat$ for each quasar epoch, respectively. 
Each value of $\sfvar$ is plotted at the median $\deltat$ in the considered bin. 
The dashed line represents a linear fit of $\sfvar$, which is parametrised as: 
\begin{equation}
\label{sfvar}
\sfvar = (0.039\pm0.004) \log \deltat + (-0.116\pm0.040).
\end{equation}
The level of fractional variation on timescales longer than a week ($\deltat\geq6\times10^5$ s) is $\sim$20--30\%, which is also in agreement with the value we have estimated from the analysis of the off-axis source position discussed above. Interestingly, if we consider a sub-sample of AGN where the count rates are computed at progressively smaller off-axis values, the comparison between two observations significantly becomes much tighter, and the amplitude of the fractional variation is proportionally reduced as well. 
With an off-axis angle $< 10^{\prime}$ (145 objects), the dispersion along the one-to-one relation reduces to 0.21 dex, with $\sfvar$ ranging between 10\%--25\% for $\deltat$ in the interval $10^6-10^8$s. With an off-axis angle $<5^{\prime}$ (48 objects), for instance, the dispersion along the one-to-one relation reduces to 0.15 dex, but the sample size is too small to carry out a statistical analysis of $\sfvar$.

Our preliminary test suggests that, when estimating variability indicators, a non negligible fraction ($\sim5-10\%$) of the quoted variability amplitude could be due to instrumental effects rather than intrinsic AGN intensity changes.  
% I should consider all observations instead of the first two...

\section{Conclusions}\label{conclusions}
The non-linear $\lx$--$\lo$ correlation observed in quasars indicates that there is a good coupling between the disc, emitting the primary UV radiation, and the hot-electron corona, emitting the X-rays \citep[e.g.][and references therein]{lusso2017,lussorisaliti2017}. 
In our previous works we established that the scatter of this relation is very small ($\sim0.2$ dex), once we take into account the combination of measurement uncertainties, variability, and intrinsic dispersion due to differences in the AGN physical properties. 
In the present manuscript, we have focussed on the various contributions to the {\it observed} X-ray variability for a sample of 791 unique quasars selected from SDSS--DR7 with X-ray data from the 3XMM--DR7 source catalogue, 250 of which have 2 or more {\it XMM} observations.

The latter objects typically vary with a standard deviation of fractional variation of 15--30\% in a time-frame $\deltat$ within the range $4\times10^6-10^8$ s. Yet, when the count rates are computed at progressively smaller off-axis angles, the same variability indicator ranges between roughly 10--25\%, suggesting that a fraction of the quoted variability amplitude could be due to instrumental/calibration issues rather than {\it true} variations in the quasar emission.

Had X-ray variability for a given set of AGN physical properties a strong dependence on redshift, this could heavily affect the determination of the cosmological parameters, unless we can capture this trend in some way. As of today, our analysis does not show any change of both the slope and the dispersion of the $\lx$--$\lo$ relation once variability and contaminants are taken into account. Yet, we still cannot completely rule out a residual dependence of the relation parameters on X-ray variability, given the lack of statistically significant AGN samples (e.g. $>1000$ objects) with multiple observations. 

The deepest X-ray surveys (e.g. {\it Chandra} Deep Field South, $\sim0.1$ deg$^2$) that probe the extended temporal baseline (long-term variability) are limited by the area, poorly sampling the high-end tail of the AGN luminosity function. On the other hand, wider-area surveys (e.g. COSMOS, $\sim2$ deg$^2$, probing the short term variability) are usually too shallow, limiting the redshift coverage (mainly redshifts less than $3-4$). The combination of future all-sky and high-sensitivity X-ray observatories (e.g. {\it eROSITA}, {\it Athena}) will allow us to investigate X-ray variability over a wider range of luminosity, redshift, black hole mass, and accretion states with unprecedented sample statistics, possibly minimizing the contribution of calibrations effects in these correlations.

%\backmatter

\section*{Acknowledgments}
The author gratefully thanks the anonymous referee for her/his comments and suggestions that have improved the clarity of the paper. 
The author acknowledges support from the \fundingAgency{European Union COFUND, Durham Junior Research Fellowship} under \fundingNumber{EU grant agreement no. 609412}. 
For all catalogue correlations we have used the Virtual Observatory software TOPCAT \citep{topcat2005} available online (http://www.star.bris.ac.uk/$\sim$mbt/topcat/).
This research has made use of data obtained from the 3XMM XMM--Newton serendipitous source catalogue compiled by the 10 institutes of the XMM--Newton Survey Science Centre selected by ESA. This research made use of matplotlib, a Python library for publication quality graphics \citep{hunter2007}.

\nocite{*}% Show all bib entries - both cited and uncited; comment this line to view only cited bib entries;
\bibliography{bibl}%

\end{document}